\documentstyle[epsf]{article}
\begin{document}
\title{Search for astro-gravity correlations}

\author{V.N.Rudenko, A.V.Gusev, V.K.Kravchuk, M.P.Vinogradov.}
\maketitle
\begin{center}
{\it Sternberg Astronomical Institute, Moscow State University,\\
Universitetskii Prospect 13, 119899 Moscow, Russia }
\end{center}
\begin{abstract}
A new approach in the gravitational wave experiment is considered.
In addition to the old method of searching for coincident reactions
of two separated gravitational antennae it was proposed to seek
perturbations of the gravitational detector noise background
correlated with astrophysical events such as neutrino and gamma ray
bursts which can be relaibly registered by correspondent sensors.
A general algorithm for this approach is developed. Its efficiency
is demonstrated in reanalysis of the old data concerning the phenomenon
of neutrino-gravity correlation registered during of SN1987A explosion.
\end{abstract}

\section{JOINT SCENARIOS FOR ASTRO-GRAVITY EVENTS}

A conventional scheme of the gravitational wave experiment on
searching for stochastic bursts of gravitational radiation
from astrophysical sources supposes a registration  of coincident
reactions of two or more spatially separated gravitational detectors.
It was considered as only way to establish a global nature of  the
detected signal which probably could be a metric perturbation
associated with gravitational wave if  the detector's isolation
was good enough [1].

A realization of this scheme requires  at least two identical
gravitational antennae  located in different points of the globe
with good synchronized clocks, good communication etc. Although
this ideology is known already thirty years the coincident experiment
in automatic regime was performed only by J.Weber during his first
observation with room temperature bar detectors located in Chicago and
Maryland [2]. Later the "coincidence searching"
episodically  have been done by several groups as a rule in the form of
joint data analysis of the electronic records of both setups
{\it a posteriori} but not on line. The recent example of such procedure
with cryogenic antennae EXPLORER and ALLEGRO is presented in the paper [3].
A reason why the detection of coincidences "on line" was replaced with
analysis {\it a posteriori} is obvious. The "on line" regime (although
it's very convenient and effective) requires an additional
electro-communication equipment. Besides it could be easy realized if the
same research group would have two equivalent detectors in disposal
(like it was in "time of room temperature bar detectors") but a complication
and large cost of modern cryogenic and interferometrical set up makes it
difficult in general. In nearest future the automatic selection
of coincidences probably will be realized  with two large scale
interferometric antennae which are under construction now in
the LIGO project [4]. At present however the coincidence analysis
{\it a posteriori} is considered as the only way of investigation stipulated
by a presence of two gravitational antennue in simultaneous operation
with equivalent sensitivity.

In last years another type of gravitational wave experiment was discussed.
The idea is to search weak perturbations of the gravitational detector's
noise background correlated with some astrophysical events such as neutrino
and gamma ray bursts [5,6,7,8]. The reason of this approach  lies in
the understanding that last stages of star evolution (such as supernova
explosion, binary coalescence, collapse etc.) traditionally considered as the
gravitational burst sources have to be accompanied also by neutrino and
very likely gamma radiation. It means in general that a detection of
neutrino or gamma ray bursts by appropriate sensors defines time marks
around which one might hope to find also exitations of the gravitational
detectors. An advatage of this method  consists first of all in a remarkable
reduction of the observational time interval and second in a potential
opportunity to accumulate weak signals. The last point is especially
interesting taking into account a deficit of required sensitivity of the
gravitational detectors available at present in the world laboratories.

The theoretical presentation of the neutrino bursts produced by
collapsing stars at the end of stellar evolution is well known, see for
example [9,10,11]. According to the theory a total energy released in the
form of neutrino radiation of all flavors has the order of value
$ 0.1 M_{\odot}c^{2} $ and a time scale of several seconds (2--20 s)
This radiation can be detected (mainly due to the inverse
$\beta$-decay reaction)  if a source is located not too far from the
Earth (10--100) kpc.
Correspondent experimental programms ("Supernova Watcher") are accepted
and carried out by the all neutrino groups having appropriate liquid
scintillation detectors [12,13] or water cherenkov detectors [14,15].
Moreover the first registration of neutrino flux from supernova as it
believes was fixed during of SN1987A explosion [16-19].
All this programms are orientated on the search  of collapsing stars in the
Galaxy and close local groups i.e. expected average rate of events is
3 per 100 years [16]. It is unlikely to wait a large increasing of
penetrating power from the neutrino telescopes in nearest future.
So Super Kamiokande detector with effective mass in ten times larger
allows a detection of 150 neutrino events per year from LMC but
only one event from Andromeda [20]. It is unrealistic to relay on
a detection neutrino from supernova in the Virgo Claster (15--20 Mpc)
which considered as one of the principal sources of a signal for
gravitational detectors. Thus a search of correlations between noise
backgrounds of neutrino and gravitational wave detectors is limited
by the condition of very low event rate $(3-10) 10^{-2} {\rm y}^{-1} $ and
an opportunity of "signal-noise enhancing" through some integrating
procedure practically is absent. Although an expected amplitude of
a solitary gravitaional pulse signal might be relatively large
up to $10^{-18}$ in term of metric perturbation from a source in
the center of Galaxy.

The other astrophysical phenomenon of our interest, gamma-ray bursts,
looks more propitious although it still remains to be confused [21].
The main attractive feature of this phenomenon is a relatively
high event rate, on average one per day. The large energy emission
evaluated for some registered gamma bursts up to the $0,1 M_{\odot}c^{2}$
together with amplitude short time variations on order of $ 0,1 s$
implies to relativistic stars as burst sources.

In process of study of this phenomenon two principal scenarios have been
 considered
in respect of the gamma-ray  bursts nature. The first one suggests its
galactic
origin associated with high velocity pulsars distributed not only in the
galactic disc but also in the Halo [22].
The second scenario appeals to a cosmological picture in which
gamma bursts are produced during catastrophic processes
with relativistic stars such as collapses, binary coalescences,
supernova explosions in distant galaxies [23]. Thus the both scenarios
deal with objects that have been considered also as sources of gravitational
radiation. Galactic pulsars
could produce only very weak GW-bursts as a result of "starquakes"
with equivalent metric perturbation on the Earth of order of
$10^{-23}\div 10^{-24}$ [24] for a source in center of Galaxy.
However authors of the papers [25,26] believe that even a more close
pulsar population in vicinity 100 pc might provide an observable
rate of gamma events $\sim 5$ per month through mechanizm of "starquake".
Then a correspondent GW burst amplitude would be awaited on the
level of $10^{-21}\div 10^{-22}$.
In the cosmological picture, if one includes into
consideration binaries with back hole components the astrophysical
forecast gives  the GW-burst event rate up to 30 per year at a metric
amplitude level of $10^{-21}$ in the solar vicinity of
50--100 Mpc [27,28].
This estimation was found supposing that only $10^{-4}$ part of
stellar rest mass energy could be converted into gravitational radiation.
A more optimistic value of the convertion coefficient $10^{-2}$ used
in the other papers [29,30] would increase the expected metric
amplitude up to $10^{-20}$.

The recent results obtained with BeppoSAX satellite and Keck II telescope
permitted to confront the gamma-ray burst GRB971214 with a galaxy
having the redshift  of $ z=3.4$. The other case is the burst GRB970508
with an optical counterpart at $z\ge 0.835$ [31]. That is the strong evidence
of the cosmological nature at least for a part of the registered bursts.
Along with these very far sources (1-10) Gpc. more close events
were registered. For example the burst GRB980425 probably was associated
with an optical object type of supernova explosion at the distance
40 Mpc ($z=0.08$) [32]. It is not completely clear how the gamma
radiation could penetrate through envelope of supernova,
how the black hole coalescence could release the gamma burst,
but the energetic of observable events definitely requires scenarios
with a crash of relativistic stars and therefore an expectation of the
gravitational radiation accompaniment seems reasonable. Moreover the
energetic estimation of the GRB971214 burst $\sim 2\cdot10^{53}$ erg even
exceeds a conventional
theoretical electromagnetic energy release $10^{51}$ erg for supernova or
neutron star binary merging [33]. It makes the models of black hole
binary mergers or rapidly rotating massive black hole with accretion,
so called "hypernova" [34], more attractive and at the same time they are
more promissing in respect of the gavitational wave output.

Thus there are serious theoretical prerequisites to search for gravitational
bursts around time marks defined by correspondent events of neutrino and
gamma-ray detectors. Now lists of desirable events can be provided
by the four world neutrino telescops and cosmic CGRO (BATSE) and BepoSAX
satellites. In this situation the key question is a sensitivity of the
gravitational detectors which are in operation at present. In fact this
is only supercryogenic resonance detector "NAUTILUS" (INFN,Frascati)
and  similary set up "AURIGA" (INFN,Legnaro) [58]
could achieve the sensitivity level $10^{-21}$ for short bursts
 $\sim 10^{-3}$ s [35].
The two cryogenic detectors mentioned above "ALLEGRO" [59] and "EXPLORER" have
the short burst sensitivity $6 \cdot 10^{-19}$ i.e. of 2,5 orders less the
desirable value. However it worth to note here that for more long signals
the estimation of its sensitivity must be increased up to $10^{-21}$ for burst
duration close to $1 sec$ due to accumulation of signal cycles (see detailes
in [36]).

Generally an improvement of detection sensitivity  depends on our knowledge
of the signal structure, arrival time etc. In this sense a theory does not
provide us a large assortiment of models for gravitational signal. Mostly
its energetic part might be presented by a short pulse with several cycles
of carrier frequency $(10^{2}-10^{3})$ Hz [24]. There is a deficit of models
with joint description of the gravitational, neutrino, and gamma radiation
output. Some examples one can find in the papers [24,29,30,37] where multi-stage
scenarios of gravitational collapse were considered in  the processes of
neutron star formation and star remnants coalescence. In such approach a
packet of the neutrino pulses separated by time intervals from few seconds
up to several days accompanied by gravitational bursts was predicted with a
total energy release up to one percent of the rest mass. The multi-stage
scenario is also typical for collapse of massive star with large initial
angular momentum [24]. A radial matter compression there might be interrupted
by repulsing bounces, fragmentation, fragments mergers or ejection of one of
them etc. In principle each of these stage could produce gravitational,
electromagnetic and neutrino bursts but a detaile description of such models
has not yet been developed.
Entirely inspite of obvious uncertainty of joint scenarios and unknown event
rate of complex collapses in the Universe an expectation of the multi-pulse
structure for a gravitational signal associated with a packet of neutrino
and gamma ray bursts is enough grounded at present.

The argumentation above stimulates one to define an optimal
data processing of the gravitational detector output
in parallel with a record of astrophysical events registered by neutrino or
gamma ray observatories. A simple comparison with an attempt to find
coincidences is insufficient due to an inevitable unknown time delay between
events of different nature [57] but mainly due to a deficit of gravitational and
neutrino detector sensitivity. Partly for this reason the attepmts of
searching for correlation between neutrino-gamma data [38] and gamma-gravity
data [39]  were not successful. It has to be done according to the optimal
filtration theory taking into account all available information concerning of
noise background and conceivable model of signal [40].

The goal of this paper is to formulate some optimal algorithm of searching
for a correlation of neutrino as well as gamma-ray events with perturbations
of gravitational bar detector. We consider this case because a noise statistics
of the resonance bar detector at present can be defined more accurately
than the statistics of a free mass gravitational antenna.

The associated goal is to apply the optimal algorithm to old data
concerning the neutrino-gravity correlation phenomen registered during
of SN1987A supernova explosion. The results reported in the series
papers [41-44] have not found any clear astrophysical explanation ,
have met some objections, and up to now continue to be subject for discussion.

\section{MLP-detection algorithm for incoherent packets of GW pulses}

In a general frame of the filtration theory it is neccessary to define
principal properties of expected signal and noise background in order to
find an optimal filtering procedure. Following the argumentation above we
take as a signal model so called "incoherent packet of pulses" in which
gravitational
events are given by an irregulary group of short impulses with two
principal parameters: arrival times $\tau_{i}$ and amplitudes $ a_{i}$.
Intervals between individual impulses might be variated in  wide
limits according to the rate of astrophysical events.
The form of individual pulse is ignored besides its duration $\hat\tau$
which is supposed to be enough short  i.e. it contains only a few periods
of the carrier frequency $\omega$ so that $\omega \hat\tau \sim 1 $ and
$\omega \sim (\omega_{0}\mp 1/\hat\tau)$ where $\omega_{0}$ is a central
frequency  of the reciever bandwidth.

A stochastic background is defined by the noises of
gravitational bar antenna. The structure of modern cryogenic antenna
contains of a cooled bar-detector, electromechanical
transducer as a read out, amplifier and a preliminary filtration link
with limited bandwidth $\Delta\omega \le \hat\tau^{-1}$ : a differential cell,
Winer-Kolmogorov filter etc. The priciple point is that one can take the
Gaussian model of output antenna noise as a good approximation having
in the mind a perfect acoustical, seismic and electrical isolation of modern
cryogenic set ups.

After these remarks one can give a mathematical formulation of the
optimal detection procedure. The antenna output $x(t)$ is an additive mixture
of the noise $\xi(t)$ and signal $S(t)$ where the last one might be described
by incoherent sequence of short "gravitational" bursts $s_{k}$ so that
\begin{equation}
x(t)=\lambda S(t) + \xi(t) \,\,\,\, S(t)=\sum_{k} s_{k}(t).
\label{exp1}
\end{equation}
Here $\xi(t)$ is supposed to be a stationary gaussian noise with the
spectral density $W(\omega)$ defined by the antenna structure;
$\lambda=(1,0)$ is a formal parameter marking a presence or absence
of the signal.  The individual pulse signal in the $S(t)$ sequence
can be presented in the complex space as
\begin{eqnarray}
s_{k}(t)&=&{\rm Re}[\tilde s_{k}(t) e^{j\omega_{0} t}],\nonumber\\
\tilde s_{k}(t)&=&
a_{k}\tilde H(t-t_{k}) e^{j(\Theta_{k}-\omega_{0} t_{k})}.
\label{exp2}
\end{eqnarray}

The new notations in these expressions $\tilde s_{k}(t)$ and $\tilde H(t)$
are  complex overlopes of the "gravitational" bursts and impulse
characteristics (Green function) of the linear antenna track
$$
H(t)={\rm Re}[\tilde H(t) e^{j\omega_{0} t}]=H_{0}(t) \cos {[\omega_{0}t +\psi(t)]}
$$
with $\omega_{0}$ as a resonance frequency of the bar antenna.

In addition to the mentioned signal parameters, --- pulse amplitudes
and arrival times, the expression (\ref{exp2}) containts also the third
parameter, --- initial phases $\Theta_{k}$. Of course an optimal data
processing algorithm depends on {\it apriori} suppositions concerning
these values.

The amplitude parameter $a_{k}$, if it is small, does not
produce any remarkable influence on the structure of data processing
algorithm. In contrast the two other parameters, initial phase and
pulse arrival time essentially affect on the optimal detection procedure.
In particulary a principal specifics of the "astro-gravity correlation
hypothesis " should be expressed in the {\it apriori} supposition that
arrival times of gravitational bursts are  located in vicinity of
astrophysical event times registered by  some independent way, i.e.
\begin{equation}
 t_{k}=t_{ak}+\tau, \quad k=[1,2...n].
\label{exp3}
\end{equation}
Here $\tau_{ak}$ are the time-marks of "astrophysical events" a total number
of which was $n$ on the observational interval $[0,T]$; \, $\tau$ is an
unknown shift between "astrophysical" and "gravitational" events.
 Admissible values of this shift have to be limited {\it apriori}  by some
interval
$(\tau_{min}, \tau_{max})$  which must be defined specially.

The problem of optimal data processing algorithm for a detecting of packet
of GW-pulses on the output of gravitational bar antnna can be solved
in the frame of Maximum Likelihood Principal. According to MLP one
has to construct a special variable, some function of antenna output
process $x(t)$,
maximization of which can provide a maximum probability to register
a signal {\it aposteriori} , i.e. refering only to the factual realization
$x(t)$ on the observational time interval $[0,T]$ and having in the mind
an available {\it apriori} information .

In the case of a signal on the gaussian noise the answer is well known:
MLP-variable $z$ is proportional to the logorithm of likelihood ratio
functional $\Lambda[x]$ [40,45]
\begin{equation}
\Lambda[x]=\left< exp\left[\int_{0}^{T}x(t)u(t)\,dt \,-\,\frac12\int_{0}^{T}S(t)u(t)\right]\right>
\label{exp4}
\end{equation}
where the reference function $u(t)$ is a solution of the integral equation

\begin{equation}
\int_{0}^{T}K_{\xi}(t-\tau)u(\tau)\,d\tau =S(t)\,\, , 0<t<T
\label{exp5}
\end{equation}
with $K_{\xi}(t)$ --- the correlation function of the $\xi(t)$ process; the
simbol $<..>$ means a statistical averaging.

In this point we must introduce  one more hypothesis concerning on {\it
apriori} signal information: namely we suppose that pulses in our packet
are rare enough and can not recover each other in time. It completely
corresponds to astrophysical expectation of small events rate for catastrofic
phenomena with relativistic stars. This hypothesis of "unrecovering pulses"
immedeately leads to the factorization of likelihood ratio functional
\begin{equation}
\Lambda[x]=\prod_{k=1}^{n} \Lambda_{k}[x]
\label{exp6}
\end{equation}
where $\Lambda_{k}$ is the likelihood ratio functional for individual
k-pulse. It obeys to the formulae (\ref{exp4},\ref{exp5}) with substitution
$u(t) \to u_{k}(t)$ and $S(t) \to s_{k}(t)$. Thus finally the MLP-variable
can be presented in the form

\begin{equation}
Z=\sum_{k=1}^{n} \ln\Lambda_{k}[x]=\sum_{k=1}^{n}z_{k},\quad
z_{k}=\ln\Lambda_{k}[x].
\label{exp7}
\end{equation}

The equations (\ref{exp4}), (\ref{exp5}), written for the individual
pulse $s_{k}(t)$  and formula (\ref{exp7}) represent a general solution
of the problem MLP-variable for the incoherent packet  of signal pulses
on the gaussian noise background. To reduce it on a practical level  one has
to find a manifest form of the reference function $u_{k}(t)$ and then to
calculate the value $z_{k}$. Below we give some approach to such procedure.

Under the natural conditions that a correlation time of the bar antenna
noise as well as a signal pulse duration are much less the observational
interval  $T$ one can expand the upper limit of integrand in the equation
(\ref{exp5}) to infinity; then a spectral thransformation of this expansion
leads to

\begin{equation}
u_{k}(\omega)\simeq s_{k}(\omega)/N_{\xi}(\omega)
\label{exp8}
\end{equation}
where the correspondent Fourier transformants are introduced:
$u_{k}(\omega)\leftrightarrow u_{k}(t)$, \,\,\, $ s_{k}(\omega) \leftrightarrow
s_{k}(t)$ and $N_{\xi}(\omega) \leftrightarrow K_{\xi}(t)$.

Then having in a mind the Parseval identity
$$
 \int_{-\infty}^{\infty}
a(t)b(t)\,dt=(1/2\pi)\int_{-\infty}^{\infty}a(\omega)b^{*}(\omega)\,d\omega
$$
one can rewrite (\ref{exp4}) for individual pulse in the following form
\begin{equation}
\Lambda_{k}[x]\simeq
\left<\exp\left[{\rm Re}\left\{\frac{1}{2\pi}\int_{-\infty}^{\infty} \frac{x(\omega+\omega_{0})
\tilde s^{*}_{k}(\omega)}{N_{\xi}(\omega+\omega_{0})}\,d\omega
-\frac{1}{8\pi}\int_{-\infty}^{\infty}
\frac{|\tilde s^{*}_{k}(\omega)|^{2}}{N_{\xi}(\omega+\omega_{0})} d\omega \right\} \right] \right>
\label{exp9}
\end{equation}

Here $\tilde s_{k}(\omega)$ is the transformant of the complex overlope
of the pulse $\tilde s_{k}(t)$. After substitution  this overlope
from (\ref{exp2}) in (\ref{exp9}) the last one is reduced to

\begin{eqnarray}{}
\Lambda_{k}[x]\simeq
\left<\exp\left[a_{k}{\rm Re}\left\{\frac{e^{-j\chi_{k}}}{2\pi}
\int_{-\infty}^{\infty}\frac{x(\omega+\omega_{0})\tilde H^{*}(\omega)}{N_{\xi}
(\omega +\omega_{0})} e^{j\omega t_{k}}\,d\omega
\right.\right.\right.
\nonumber\\
\left.\left.\left.
-a^{2}_{k}\frac{1}{8\pi}\int_{-\infty}^{\infty} \frac{|\tilde H(\omega)|^{2}}
{N_{\xi}(\omega+\omega_{0})}d\omega\right\}\right]\right>
\label{exp10}
\end{eqnarray}
with $\tilde H(\omega) \leftrightarrow \tilde H(t)$ and $\,\chi_{k}=
\omega_{0}t_{k} -\Theta_{k}$.

It is convenient to present the expression (\ref{exp10}) in terms of
the output antenna variable $\tilde y(t)$ which  is a result of passing the
input variable $x(t)$  (\ref{exp1}) through some optimal "data
processing filter" with transfer function $ K_{opt}=[H^{*}(\omega)/
N_{\xi}(\omega)]\exp(-j\omega t_{0})$, where $t_{0}$ is a filter time
delay. Then (\ref{exp10}) can be converted into
\begin{equation}
\Lambda_{k}[x]\simeq \left<\exp\left\{a_{k}{\rm Re}\left[e^{j\psi_{k}}\tilde y(t_{k})\right]-
a_{k}^{2}\sigma^{2}/2 \right\}\right>
\label{exp11}
\end{equation}
with notations: $\psi_{k}= \omega_{0}(t_{0}+t_{k})-\Theta_{k}$ and
$\sigma^{2}=(1/\pi)\int_{-\infty}^{\infty} |K_{opt}(\omega)|^{2}
/N_{\xi}(\omega)\,d\omega$ --- the output noise dispersion.

Finally, introducing an amplitude of the output signal reaction
$A_{k}=|<\tilde y(t_{k})>|=a_{k}\sigma^{2}$ one comes to the following
expression for the k-likelihood ratio
\begin{equation}
\Lambda_{k}[x]\simeq \left<\exp\left\{(A_{k}/\sigma^{2}){\rm Re}\left[e^{j\psi_{k}}\tilde y(t_{k})\right]
-A_{k}^{2}/2\sigma^{2}\right\}\right>
\label{exp12}
\end{equation}

The formula (\ref{exp12}) gives in principle an answer to the question about
the structure of MLP-variable, but  it contains signal pulse parameters
$A_{k},\Theta_{k},t_{k}$ which are unknown {\it apriori}. To avoid this
problem one can use a so called "generalized form of MLP" [45] when unknown
parameters are replaced by their "maximum likelihood evaluations"
$\hat A_{k}, \hat \Theta_{k}, \hat t_{k}$ which can be taken as
solutions of the following extremum equations

\begin{equation}
\partial z_{k}/\partial A_{k}=0,\,\,\partial z_{k}/\partial \Theta_{k}=0,\,\,
\partial z_{k}/\partial t_{k}=0
\label{exp13}
\end{equation}
A direct calculation  with $z_{k}=\ln \Lambda_{k}[x]$ from (\ref{exp12})
leads to conclusions that \\
a) the MLP-evaluation of the amplitude coincides with the overlope
of a narrow bandwidth process on the antenna output $R(t)$
\begin{equation}
\hat A_{k}^{2}=\{Re[e^{j\hat\psi_{k}} \tilde y(t_{k})]\}^{2}=|y(t_{k})|^{2}=
R^{2}(t_{k})
\label{exp14}
\end{equation}
and then a recipe for construction of MLP-variable is
\begin{equation}
z_{k}=(\hat A_{k}^{2}/2\sigma^{2})=(R^{2}(t_{k})/2\sigma^{2})
\label{exp15}
\end{equation}
b) the MLP-evaluation of the unknown time shift $\tau$ between the
"astrophysical" time mark $\tau_{ak}$ and arrival time of "gravitational"
signal $\tau_{k}$ is defined by a position of maximum of the function
$z_{k}(\tau_{ak}+\tau)$ in the $\tau$ space.

The conclusions above correspond to the supposition that parameters
of signal pulse are definite but unknown values. There is also other
conceivable case when one considers initial signal phases $\Theta_{k}$
as stochastic variables with uniform distribution in the interval $[0,2\pi]$.
Then after a statistical averaging one can find  a different form of the
MLP-variable
\begin{equation}
<\Lambda_{k}[x]>=\exp\left[-\frac{\hat A_{k}^{2}}{2\sigma^{2}}\right]
I_{0}\left(\frac{\hat A_{k}R(t_{k})}{\sigma^{2}}\right)
\label{exp16}
\end{equation}
with the following equation for amplitudes $\hat
A_{k}$
\begin{equation}
\hat A_{k}= R(t_{k}) \frac{I_{0}[\hat A_{k} R(t_{k})]}{I_{1}[\hat
A_{k}R(t_{k})]}
\label{exp17}
\end{equation}
$I_{0},I_{1} $ in the (\ref{exp16}),(\ref{exp17}) are the modified Bessel
functions.

A correspondent MLP-variable  for the case of stochastic initial phase
$\Theta_{k}$   looks like
\begin{equation}
z_{k}= \ln I_{0}\left( \frac{\hat A_{k} R(t_{k})}{\sigma^{2}} -
\frac{\hat A_{k}^{2}}{2\sigma^{2}}\right)
\label{exp18}
\end{equation}
A solution of  the equation (\ref{exp17}) is given on the Fig.1.
It demonstrates that a difference between estimations (\ref{exp14}) and
(\ref{exp17}) is essential only for small signals with amplitudes
$A_{k} > \sigma$. For the amplitudes $A_{k}\ge 2\sigma$ the
both estimations practically coincides and recommends to take  a value
of the output overlope $R(t_{k})$ as a MLP-evaluation of
the $A_{k}$.  Then the difference between MPL variables (\ref{exp15})
and (\ref{exp18}) also vanishes.

Now coming back to the expression (\ref{exp7}) we can summarize
the results. In the frame of the model
of incoherent packet of signal pulses on gaussian
narrowband noise background the MLP-algorithm recommends to compose
a following variable

\begin{equation}
 Z=\sum_{k=1}^{n}  (R^{2}(t_{k})/2\sigma^{2})
\label{exp19}
\end{equation}
which is the sum of quadratic counts of the antenna output overlope
taken in times of astrophysical events with some small shift
$\tau \,\,$ (\ref{exp3}); the sum is accumulated on the interval of
observation which {\it aposteriori} contained $n$ events.

\begin{figure}
\epsfxsize=\textwidth
\epsfbox{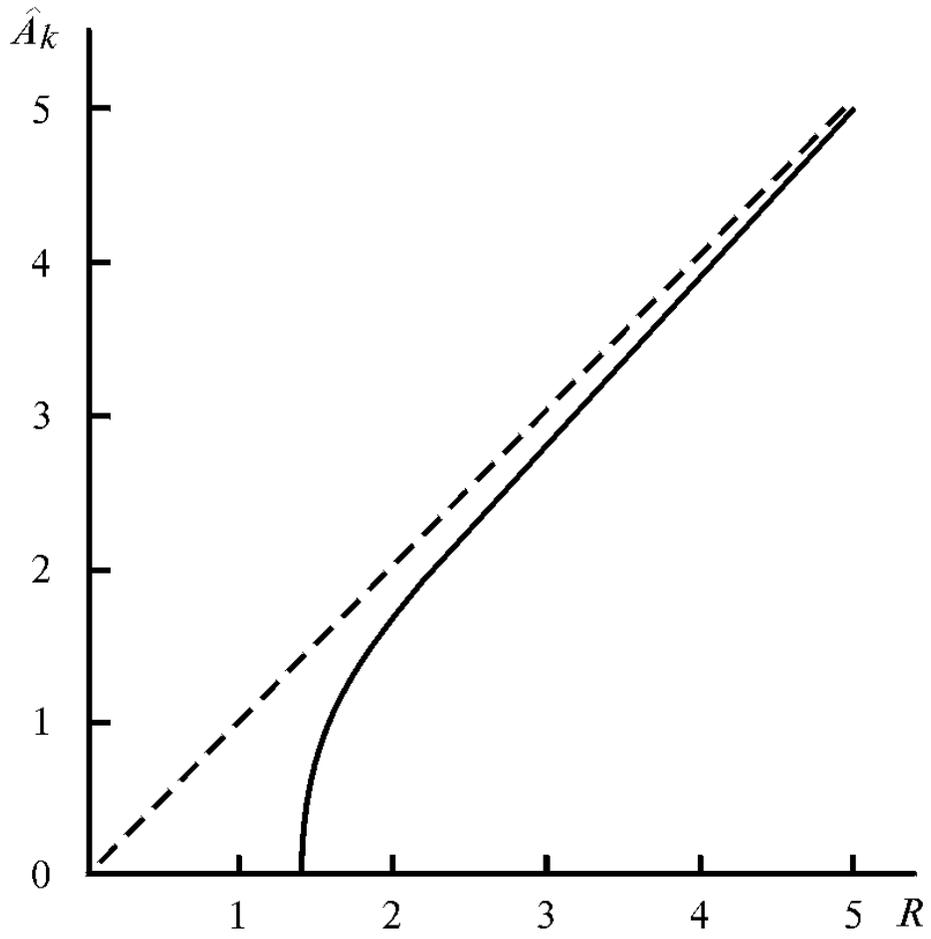}
\caption{MLP-estimation of the signal amplitude}
\end{figure}

\medskip

Then it is recommended to find an absolute maximum of $Z$ through
variations of the shift $\tau$ (see (\ref{exp13}) and point $b)$), i.e.
to get over a new so called absolute maximum --- variable
\begin{equation}
Z_{max}= \max_{\tau} Z(\tau), \,\,\,\, \tau \in [\tau_{min},\tau_{max}]
\label{exp20}
\end{equation}
A value of $\tau_{opt}$ which provides a maximum of $Z(\tau)$ should
be taken as MLP-evaluation of the real time shift between astrophysical
event and gravitational signal (in our simple approach the shift is
supposed to be the same for all events, --- a hypothesis of "homogenity
of events"). As we remarked already there is no  a definition of the
$\tau$-interval limits inside of the statistical model; it has to be choosed
on a base of additional physical arguments.

A strategy of the operator performing a data processing of gravitational
antenna output and having a list of "astrophysical events"
can be thought in the frame of Neuman-Pirson approach (under condition
of {\it appriori information deficit}). After composing the $Z_{am}$
variable one has to compare it with {\it a threshold} defined by statistical
properties of $Z_{max}$. A crossing of the threshold  would mean "a presence
of signal" with an accuracy of the "false alarm" error.
Thus such strategy supposes a preliminary knowledge of $Z_{max}$
statistics. It might be taken from theoretical suppositions or to
be a result of some empirical study of the output antenna realization.

\section{STATISTICAL PROPERTIES OF THE MEASURABLE VARIABLES}

There are three  observant variables involved in the MLP data processing.
These are  the squared overlope of the antenna output $ R^{2}(t)$
(\ref{exp14}), the sum of overlope counts $Z$ (\ref{exp19}), taken in
the moments of astrophysical events on the observational interval $[0,T]$
and the maximum value of this sum $Z_{max}$ (\ref{exp20})
corresponded to the optimal time shift.
Statistics all of them can be calculated analitically
if we accept the gaussian approximation of the bar antenna noise.
As experiment has shown this supposition is very close to reality with an
exception of large energetic thresholds where the thermal statistics can
be distored by stochastic nongravitational hindrances (a correction which
could be introduced in this case  we discuss in section 5).

The formulae (\ref{exp19}),(\ref{exp20}) were derived in dimentionless
form. For a comparison with experiments it is useful to have also
dimentional  expressions for these variables in kelvin degrees.

\subsection{Statistics of the squeared overlope}

It is well known that thermal oscillations of the resonance bar detector
are described  as a narrow band gaussian stochastic process
$x(t)=A(t)\cos\omega_{0}t - B(t)\sin\omega_{0}t$ with slow changing quadratures
$A(t),B(t)$ having the correlation function $k(\tau)=\sigma_{0}^{2}
\exp(-\gamma
|\tau|)$, where $\sigma_{0}^{2}=kT_{0}/m\omega_{0}^{2}$ is the brownian
dispersion
and $\gamma=\omega_{0}/2Q$ is the relaxation index ($m,T_{0},Q$ are the bar
equivalent mass, absolute temperature and quality factor).

After preliminary filtration (a differential link, Winer-Kolmogorov filter
etc.) $x(t)$ reduces to some narrowband process inside a limited
bandwidth $\Delta \omega$ with squeared overlope $R^{2}=(\Delta A)^{2} +
(\Delta B)^{2}$. This value is proportional to an {\it "energy innovation"}
(or variation of energy) of the bar $E(t)$ during the time
$\Delta t =\Delta\omega^{-1} \ll \gamma^{-1}$.
$$
E(t)=m\omega_{0}^{2}R^{2}(t)/2,\,\,\, <E(t)>=kT_{0}2\gamma \Delta t.
$$
 Correspondent variation of the quadrature $\Delta A$,or $
 \Delta B$ has a correlation function type of
 $k_{\Delta}(\tau)=\sigma^{2}\rho (\tau)$ where
$$
\left\{
\begin{array}{lcr}
\rho=1-(|\tau|/\Delta t),\,\,|\tau|\le\Delta t  \\
\rho=0,\,\,|\tau|>\Delta t. \\
\end {array}
\right.
$$
and the value of dispersion $\sigma^{2}$ is coupled with the
brownian dispersion $\sigma_{0}^{2}$ through an effective noise temperature
$T_{e}$ of the bar
$$
\sigma^{2}=(kT_{e}/m\omega_{0}^{2})=\sigma_{0}^{2}(T_{e}/T_{0}),\,
\,\, T_{e}=T_{0}(2\gamma\Delta t).
$$
The correlation function of the squeared output overlope can be easy
calculated in the usual form $K(\tau)=<R^{2}(t)R^{2}(t+\tau)>-<R^{2}(t)>^{2}$
which results in

\begin{equation}
K(\tau)=4\sigma^{4}\rho^{2}(\tau)
\label{exp21}
\end{equation}
The formulae above  show that the correlation coefficient of the
squeared antenna overlope $\rho^{2}(\tau)$, falls down to zero at the
"innovation time scale" $\Delta t$. Thus independent counts under a discrete
presentation of the output overlope $R(t) \to R(t_{k})$ must be separated
by time distances
$(t_{k+1}-t_{k}) \ge \Delta t $.

\subsection{Statistics of the sum of overlope counts}

The next variable of our interest is the sum of counts of output overlope
taken in
the times of astrophysical events (\ref{exp19}). It is convenient to
normalize this variable dividing it on a total number of events on
the observational interval $[0,T]$. Then the new variable $C=Z/n$ will
be proportional to {\it a "selected mean value"} of energy innovation
$$
\bar E = (1/n) \sum_{k=1}^{n} E(t_{k})
$$
collected on the observational interval in the special time marks ---
astrophysical events so as
\begin{eqnarray}
C=Z/n&=&(1/n)\sum_{k=1}^{n} R^{2}(t_{k})/2\sigma^{2}=
\nonumber\\
&=&(1/nkT_{e})\sum_{k=1}^{n} E(t_{k})=(1/kT_{e})\bar E
\label{exp22}
\end{eqnarray}
If the number of events in the sum (\ref{exp22}) is larger the thirty
a distribution of the C-variable assimptotically should be the gaussian one
according to the "central limiting theorem" with  the mean value
$<C(t)>=1$ or $ <\bar E>=kT_{e}$. The correlation function
$ \rho_{c}=<C(t_{1},t_{2}...t_{n})\,C(t_{1}+\tau,t_{2}+\tau,
...t_{n}+\tau)>-<C(t_{1},t_{2}...t_{n})>^{2}>$ has a structure
$$
\rho_{c}(\tau)=\frac{1}{n^{2}}\left[n\rho^{2}(\tau)+\sum_{i=1}^{n}\sum_{k=1,k\ne i}^{n}
\rho^{2}(t_{i}-t_{k}+\tau)\right]
$$
which demonstrates the presence of a principal peak in the region $0\le
\tau\le \Delta t$ with a parabpolic degeneration in time and a series peaks in
the points where $\tau=(t_{i}-t_{k})$. Such nontrivial structure produces
some pecularity in a definition and calculation of the "correlation time"
for C-variable. Here we  would like only remark that under a supposition
that the sequence of astrophysical events is a poissonian flux of pulses,
the expression of $\rho_{c}(\tau)$ can be reduced to the following form
(where $\Delta t \le \tau \le T $)
\begin{equation}
\rho_{c}(\tau)=\frac{1}{n}\left[\rho^{2}(\tau)
+\frac{1}{\pi}(n-1)\left(\frac{\Delta t}{T}\right)\left(1-\frac{|\tau|}{T}\right)\right]
\label{exp23}
\end{equation}

Under reasonable suppositions that the total number of events $n$ on the
observational interval $T$ is not too large $n(\Delta t/T) \ll 1$ and the
correlation time of $C(\tau)$ is limited $|\tau_{c}|\ll T$ the formula
(\ref{exp23}) might be simlified
\begin{equation}
\rho_{c}(\tau) \simeq \rho^{2}(\tau)/n,\,\, \longrightarrow \,\,
K_{\bar E}(\tau)= \rho^{2}(\tau)\,(kT)^{2}/n
\label{exp24}
\end{equation}
A dispersion of the $C$-variable (and $\bar E$) is depressed in the factor
of $(1/n)$
in compare with the $R^{2}$-variable in agreement with the statistical
property of the sum of identical independent counts.

\subsection{Statistics of the  absolute maximum of C-variable}

As above we will consider the normalized sum of the overlope counts
i.e.instead of $Z_{max}$ (\ref{exp20}) one deals with
$C_{max}=\max_{\tau}Z(\tau)/n$.
The maximum has to be found through time shift variations on the
apriori given time interval (\ref{exp20}). Let's accept that the time
shifts are produced by descrete steps $\delta t$. Then  we have the
output combination of values $\{C(t_{ak}+m\delta t)\}$,
($m=1,2...L$) with a total
number $L=(\tau_{max}-\tau_{min})/\delta t$.

In the case of gaussian statistics of the $C$-variable a solution for its
absolute maximum distribution might
be taken from literature. In particulary one can use the Cramer formula [46]
which presents the absolute maximum statistics $C_{max}$ through another
auxillary stochastic parameter $\xi$.
\begin{eqnarray}
C_{max}&-&<C>\simeq \nonumber\\
&&\simeq\sqrt{\frac{1}{n}} \left[\sqrt {2\ln \mu(\Delta\tau)} + \xi/
\sqrt{2\ln\mu(\Delta\tau)}\right] \label{exp25}
\end{eqnarray}
where $\xi$ has a probability density
\begin{equation}
w(\xi)=e^{-\xi}\exp(-e^{-\xi})
\label{exp26}
\end{equation}
with a mean value $<\xi>=0,577$ and dispersion $\sigma_{\xi}^{2}=
\pi^{2}/6 $.

The formulae (\ref{exp25}),(\ref{exp26}) are true
in the assimptotical sense, i.e. under ($\Delta \tau/\delta t)\to \infty $.

The parameter $\mu$ in the formula (\ref{exp25}) depends on the region of
time shift variations $\Delta \tau = (\tau_{max}-\tau_{min})$ and  a second
derivative of the correlation coefficient of $C$-variable $ \rho^{2}(\tau) $
in the point $\tau=0$ so that

\begin{equation}
\mu(\Delta\tau)=(1/2\pi)\Delta\tau\sqrt{-2\rho''(0)}
\label{exp27}
\end{equation}
A calculation the value $\ddot R_{c}(0)$ for the processes of Markov type
is always a nontrivial procedure. In our case an estimation can be done
through the approximation of Owen functions [47] and results in
\begin{eqnarray}
\mu(\Delta\tau)&=&\frac{1}{\pi}\frac{\Delta\tau}{\delta t}
\sqrt{\frac{1-\rho^{2}}{1+\rho^{2}}}, \nonumber\\
&&\rho^{2}=(1-\delta t/\Delta\tau)^{2}.
\label{exp28}
\end{eqnarray}
The formulae (\ref{exp25}),(\ref{exp26}), (\ref{exp28} )in principle solve
the problem of calculation a "probability of chance"
to exceed some threshold level $C_{th}$
for the absolute maximum variable $C_{max}$ .

\section{NEUTRINO-GRAVITY CORRELATION EFFECT OF SN1987A}

As a test of the proposed algorithm we consider
its application to the phenomenon of "neutrino-gravity correlation effect"
reported in the series papers by the RTM-collaboration {\bf (}INFN, Univ.
"La Sapienza", "Tor Vergata" (Roma), Inst. Cosmogeofisica CNR (Torino),
Univ. Maryland (Washington) and Inst. Nuclear Res. Rus. Ac (Moscow){\bf )}
[41-44].

The effect consists in fixation of remarkable correlation
during of SN1987A phenomenon
between the unified noise background of room temperature gravitational
bar detectors in Roma and Maryland at the one side and a neutrino background
registered by the Mont Blanc neutrino scintillator at the other side.
A direct interpretation of this correlation as an affect of gravitational
and neutrino radiations from a collapsing star have met objections
from the point of view a required energy of gravitational wave.
There was a deficit of two order of value in a conventional estimation
of the gravitational radiation output from supernova at the distance of
BMC in compare with the room temperature bar detector sensitivity [41].
Later several other investigations were carried out in attempts to clarify
a nature of this effect which probability of chance was evaluated as
extremely small, order of $10^{-6}$ [42].  In that number a searching of
any correlation with other elementary particle backgrounds [48], with
seismic noise background [49] etc. Besides a dynamics of joint antenna
pattern  of gravitational detectors in Roma and Maryland was calculated
[50] and some hypothesis of a new physics also were considered (see
examples in [51].
Nevertheless it did not lead to any definit model of the phenomenon.
Then a computer simulation of the neutrino and gravity data  was
carried out  to prove that the "$\nu g$-correlation effect" could be
a usual statistical fluctuation if one would be correct in probability
of chance estimation [52]. However RTM collaboration did not accept this
critics arguing that the authors of [52] did not use the real  experimental
data and presented some contrary argumentation in favour of the objective
character of the effect [53].

In this section we present the results of our analysis of the real data
kindly provided for us by the RTM group. With it we follow the algorithm
developed in the previous sections  making a coparison with the RTM
methodics.

\subsection{Method and results of RTM group}

The bank of data containing joint records of "energy innovations"  of the
gravitational detectors in Roma and in Maryland  was limited by the time
interval from UT, 12h00m, Feb. 22 --- UT, 06h00m, Feb. 23.
At the same time interval there was a list of neutrino events corresponded
to the stochastic background counts of the  LSD  neutrino scintillator in
the program "Supernova Watcher". All data were presented in the digital
form. A sampling time of the gravitational records  $\Delta t=1$ s.
was also equal to the "innovation time" interval (i.e. a bandwidth
of the filtering tract was $\Delta\omega=1/\Delta t$).
A sampling time of the neutrino counts ( an accuracy of the event time marks)
was 0,01 s. There were no joint data after UT, 07h00m because
the Maryland detector had stoped an operation for technical reasons.

The neutrino list had a singularity in the region 2h52m, Feb. 23: there was
a group of five $\nu$-pulses with very small poissonian probability of
chance. These neutrino was detected inependently and beforehand an
information about optical observation of the supernova was received.

Side by side with traditional "coincidence methodics" RTM group have applied
an original method of analysis composing from gravitational data
an auxilary ststistics which was {\it the sum of energy innovations taken
in the neutrino time marks } normalized to the number of events. In fact
the RTM group have anticipated the optimal strategy of MLP approach resulting
in the C-variable (\ref{exp22}) as "a sufficient statistics". The reason of
such choice RTM group have seen in the physical sense of the C-variable
as a value proportional to {\it the correlation function} between two
stochastic serieses: counts of the gravitational detector energy variations
and neutrino events (in the last series only time marks of events was
essential because the amplitudes were fixed by the threshold selection).
As we mentioned in the previous section there is also another important
physical sense of the  C-variable as a mean value of "gravitational
energy innovation" calculated on the base of "neutrino times".

A generalization introduced by RTM group consisted also of the decision
to use a combined energy innovation  composed by a linear combination
of counts of two appriori independent set ups in Roma and Maryland.
This so called "a net exitation method" in therminology of the paper [52]
has a clear argumentation considering the two gravitational detectors
as links of the one united wideband gravitational antenna under a global
gravitational wave  influence (frequencies of Roma and Maryland
detectors were different so the unified antenna received an energy
from different spectral component of GW-pulse). Thus RTM group have dealt
with C-variable in the dimentional form

\begin{equation}
C(\tau)=(1/n)\sum_{k=1}^{n} [E_{R}(t_{k}+\tau)+ E_{M}(t_{k}+\tau)]
\label{exp29}
\end{equation}
which corresponds in fact $\bar E $ in our notation (\ref{exp22}), but
for a convenient comparison we will keep the RTM definitions in this section.
The summation (\ref{exp29}) was made with some normalization factor,
reflected in noise temperatures of the both detectors.

The value of $C(\tau)$ have been collected into two hours time window
according to neutrino marks inside of it. The procedure was repeated after
displacement the window on half hour along the observation time interval
and so on.
The time shift $\tau$ was taken equal --1.4~s according to the estimation
made in the first paper [41] through a digital filter applied to the Roma
detector data and repeating the structure of the "five-neutrino group"
registered by LSD at 2h52m. Later in the paper [42] this shift was
reduced to --1.2~s as a more optimal value resulting in the largest
meaning of the $C$-variable.

The main result of the RTM group analysis consists in the statement that
the $C$ statistics reaches  the maximum value $C_{exp}=72.5 K^{0}$ on the
two hours interval around  the point 2h52m where the number of
registered neutrino events was $n=96$.

To prove it RTM group have made a simulation of the neutrino events
using a generator of stochastic numbers which provided neutrino time
marks according to the poissonian low (the total number of events have
been fixed by the real neutrino list). Having this artificial "neutrino
flux" RTM group could calculate corresponded values of $C$-variable for
each two hours interval with variation of the time shift if neccessary.

A presence of the "$\nu g$-correlation" effect  was demonstrated  on
two type of graphs. The first was  a relative number of cases when
a simulated "artificial" $C(\tau=-1.2 {\rm s})$ exceeded of the observable
in experiment value $C_{exp}$ versus of consequent two hours intervals,
see Fig.~2,~a. The second was an analogical relative number but calculated
on the two hours interval around of 2h52m (the interval of "neutrino
singularity") versus of time shift  which was variated around the
stationary meaning (--1.2~s), see Fig.2~c. The both graphs have shown
an exclusivity of the experimentaly observed data: there were deep
downfalls at the place around of 2h52m, and in the point
of zero time shift at the Fig.2~a. A presence of these downfalls
means a registration of a very rare event. The RTM group used two ways
for a probability of chance  estimation. First, a simple utilization of
the "binominal formula"
$p=m/n$, where $m$-the number of cases when $C\ge C_{exp}$ and $n$-the total
number of tests.
Second, an empirical construction of the $C$ statistics distribution
through the simulation  of the neutrino events flux (each simulated
neutrino series gave a definit value of $C$ --- a one point in the empirical
differential distribution graph). Having this distribution in disposal
one could easy evaluate  the chance probability to get a definit
$C$ value. The $C$ distribution from the paper [42] is presented
on the Fig.2~e together with position of the observable in experiment
value $C_{exp}$.

The both ways gave a chance probability $p\simeq 10^{-3}$ for the effect on
Fig.2~a and extremely small $p\simeq 10^{-6}$ for Fig.2~c. Espessially
the last fact was interpreted as the detection of abnormal correlation
between gravitational and neutrino data in the time around 2h52m UT
--- so called "neutrino-gravity correlation effect" associated with
SN1987A.

\begin{figure}
\epsfxsize=\textwidth
\epsfbox{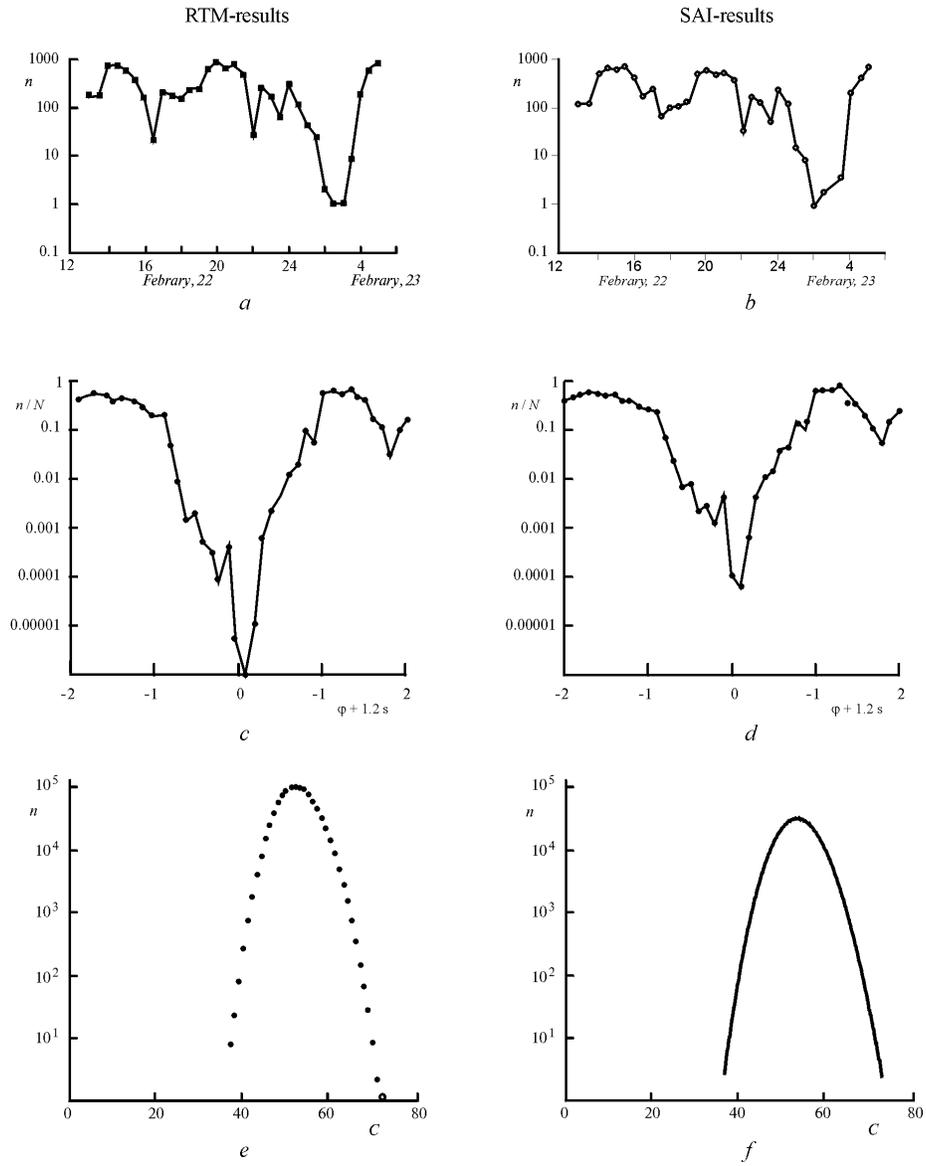}
\caption{a, b --- A time evolution of number of cases $C \ge C_{exp}$;
c, d --- Number of cases $C \ge C_{exp}$ versus of time shift on the
interval 1h52m--3h52m; e, f --- $C$-distribution under
$\nu$-set simulation}
\end{figure}

\subsection{Method and results of SAI group}

The theory of the method we used in our reanalysis was given above
in the sections II. It corresponds to the RTM method besides
the fact of replacement the $C$-variable by the $C_{max}$ statistics
but also applied to the combined R\&M data in the manner of
"net exitation" approach.

\subsubsection{Gaussian predictions}

First of all we had possibility to forecast an expected result of
our reanalysis on the base of gaussian approximation in the section
III, C. A good agreement of the experimental data (output realization
of the gravitational detectors) with hypothesis of gaussian distribution
was demonstrated more then once and in particulary in the paper [42].
The calculated and measured noise temperatures $T_{e}$
were for Roma detector $T_{e}=T_{R}=28.6 K^{0}$ and for Maryland
detector $T_{e}=T_{M}=22.1 K^{0}$ (the normalization factor
in our reanalysis was equal $\varepsilon=T_{M}/T_{R} \simeq 0.77$ which
is very close to the value of RTM group 0.75).
Then estimations of the mean value and dispersion of $C$-variable
for the "net exitation"structure and dimentional form ($C\to\bar E$)
(\ref{exp29})
according to formulae (\ref{exp21}),(\ref{exp24}) has to be
\begin{eqnarray}
k^{-1}<C>&=&(T_{R}+T_{M})\simeq 51 K^{0} \nonumber \\
k^{-1}\sqrt K_{\bar E}(0)&=&\sqrt {[(T_{R})^{2}+(T_{M})^{2}]/n}
\simeq 3,7 K^{0}, \nonumber\\
&&(n=96).
\label{exp30}
\end{eqnarray}
Thus in the region of the effect (two hours around 2h52m)
the theory forecasts the $C$-variable distribution in the gaussian
form with central point $51 K^{0}$ and effective width $3.7 K^{0}$.

As we have seen a principal statistics of MLP algorithm is the
absolute maximum of $C$ under time shift variations i.e. $C_{max}$
An expected mean value of the $C_{max}$ is given by the formula (\ref{exp25})
after its statistical averaging
\begin{eqnarray}
<C_{max}>&=&<C>+ \nonumber\\
&+&\sqrt {K_{\bar E}(0)}\,\left[\sqrt{2\ln\mu} + <\xi>/\sqrt{2\ln\mu}\right]
\label{exp31}
\end{eqnarray}
It was mentioned that $<\xi>=0.577$. In the $\mu$ estimation a principal
role belongs to the range $\Delta \tau$ and step $\delta t$ of time shifts.
There is no recommendation for a choice of them inside the MLP algorithm.
It has to be done on physical arguments. In our reanalysis it was taken
$\Delta \tau =\mp 100 {\rm s}$ and $\delta t=0.01$ s corresponding to the
experimental data specifics.Then the formula (\ref{exp28}) gives
$ \rho^{2}\simeq 0.1$ and $\mu\simeq 6.46 $. A substituition of these numbers
into (\ref{exp31})  results in the evaluation of mean value $<C_{max}>$:
in the region of the effect $<C_{max}>\simeq 65 K^{0}$ (if $k=1$).

The formulae (\ref{exp25}--\ref{exp28}) permit to estimate also a width
of $ C_{max} $-distribution as well as its form and then to find a "false
alarm error" or "chance probability" for any value of $C$ realized in
experiment. However due to dependence of these estimations on a choice
of characteristic times  we do not do it here but instead
we present below results of our empirical data analysis in the manner
similary to RTM group.

\subsubsection{Empirical analysis}

In principle an empirical analisis has a conventional advantage of refusing
from any hypothesis {\it apriori} on respect with a distribution
low of the data under consideration.  At the same time a task  of
reconstruction statistical properties of observable variables on the
base of only one unique realization of the stochastic process belongs
to the family of "ill posed" problems and uncertaities of reconstruction
might be large enough to make this method ineffective. So each step
in empirical data analysis must be estimated in respect of possible errors.

The procedure of MLP algorithm factually is a very delicate filtration
process in attempt of detecting a weak signal strongly covered by the noise.
As illustration of this idea one can look at the output realization of Roma
gravitational detector during the time Feb. 22--23 (Fig. 3) (a computer
reconstruction of the digital data). It is unlikely to extract a signal
from this background without special very sophysticated recipes.

\begin{figure}
\epsfxsize=\textwidth
\epsfbox{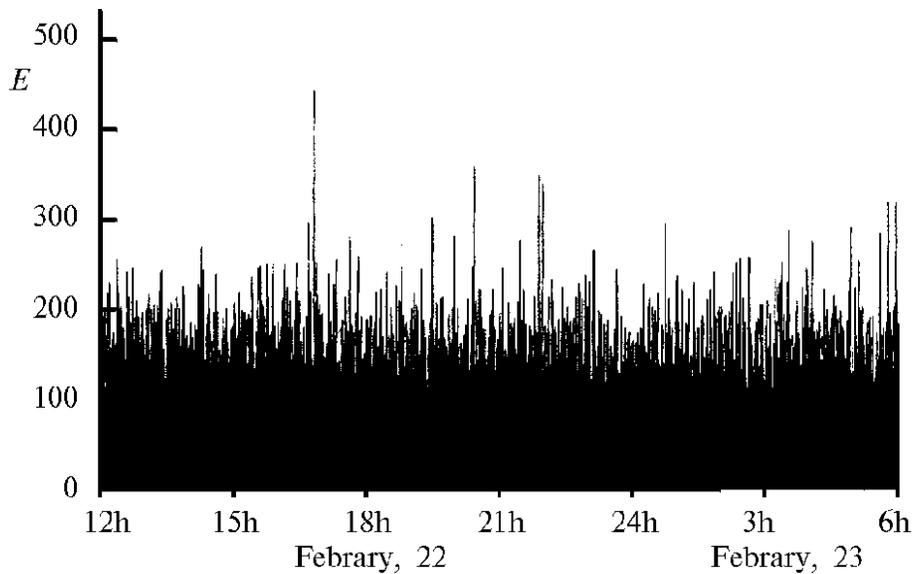}
\caption{Output record of the Roma detector from 12h 22 Feb up to 6h 23 Feb.}
\end{figure}

\medskip

In our "real data" reanalysis we started with repetition of two
tests of RTM group. i) Using "neutrino two hour serieses" simulated by
poissonian generator it was checked how often an immitational
"two hour $C$-variable" taken with fixed shift $\tau=-1.2$ s exceeded
the experimental value of $C$. Our results  completely confirmed
the results of RTM group, see Fig.2 b: we have got a singularity at the
two hour interval around of 2h52m. ii) At the region of the effect
it was checked how often a "simulated $C$ with different shifts" exceeded
the experimental $C_{exp}$ with selected shift $\tau=-1.2$ s. Again our results
confirmed a presence of singularity with slightly different estimation of
the chance probability ($\simeq 10^{-5}$), see Fig.2 c.

We used for estimation of the chance probability the same "binominal
formula" $p=m/n$ as RTM group although a rightfullness of it
was criticized in the paper [52] by the refering to the "absence of
independency" between different counts of $C$ variable. However
one can show that
for enough high values of $C_{k}$ they might be considered as independent [54] .
More in detail: the number of independent counts $n^{\star}$ in the
total sample number $n$ is defined as
\begin{eqnarray}
n^{\star}=\frac{n}{1+(n-1)r},&\quad&
r=f(C/\sqrt{K_{c}(0)})R(C_{k}),\nonumber\\
f(x)&\simeq& x (\Phi)'(x).\nonumber
\end{eqnarray}
where $\Phi (x) $ is the probability integral and $R(C_{k})\simeq
n(\Delta t /T)\ll 1$ is the correlation coefficient of $C_{k}$.

For $C\ge \sqrt K_{c}(0)$,\,\,  $r(x) \to 0$, and the $n^{\star} \to n$
i.e. for relatively large values of $C$ practically all samples
are independent.

We also reconstructed the empirical $C$ distribution  at the region of
the effect and found the same graph as RTM result on the Fig.2 f
which was centered in the point $52 K^{0}$ in a good agreement with
theoretical forecast of gaussian approach ($51 K^{0}$).

Having got these confirmations we must not forget however general
properties of solution of ill posed problems: a reliability falls
down on the wings of reconstructed distribution. So we could accept
the empirical estimations of chance probability above not literally
but only on the order of value.

After all of this we can consider the key point ot our reanalysis
which is the following: the "binominal formula" and "C" distribution
are not adequate statistics for the "probability of chance" evaluation
in the expriment under consideration. The matter is the estimations
above did not take into account a selection of data through  time
shift variations. As the general MLP algorithm recommends it has to be
done with help of the {\it absolute maximum distibution} $C_{max}$.
In the empirical method a reconstruction of this distribution on the
interval of the effect goes through the following procedure:
one simulates a "neutrino series" with $n=96$ and then variates the time
shift to find an optimal one provided a maximum value of $C=C_{m}$.
This value will be the one point of the $C_{max}$ distribution.
Independent repetitions of the procedure lead to reconstruction of the
complete distribution. It is naturally to wait that the selection
will move the $C_{max}$ distribution to the region of larger values
of $C$ and increase the chance probability.

\begin{figure}
\epsfxsize=\textwidth
\epsfbox{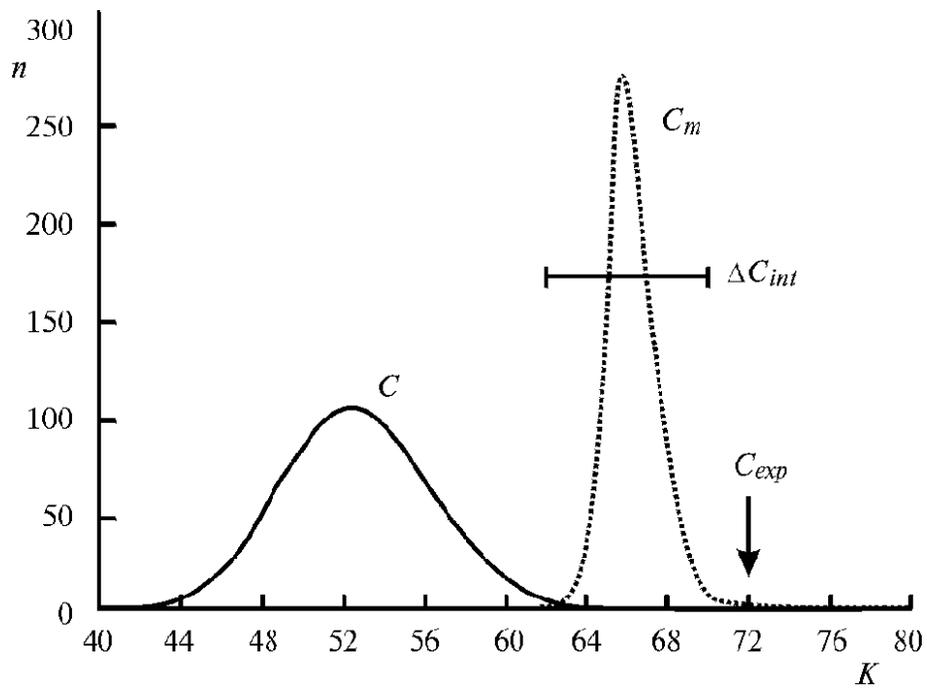}
\caption{Empirical distributions $C$ and $C_{max}$}
\end{figure}

\medskip

At the Fig. 4 we present the results of reconstruction of $C_{max}$
distribution together with $C$ distribution and mark the value of
experimentally registered effect $C_{exp}=72,5 K^{0}$. A reconstruction
of $C_{max}$ requires a much more computer time then for $C$ variable,
so the graph on Fig.4  contains $10^{3}$ points.
Two characteristic times {\it the range and step} of time
shift variations were choosed on the base of following arguments.

The range of time shift can not be too larger of an average time distance
between poissonian neutrino events so as in opposit case  a "time
shift operation" could capture edditional "neutrino" from neighbour
two hours intervals. The  average interpulse distance in the
exprimental LSD neutrino record was 70--80 s so we took the
range of time shift $\Delta \tau =\mp 100$ s.

The step of time shift was taken equal to the shortest sample time
of the data available i.e. to the "neutrino sample time" $\delta t=0,01$ s. It is clear that a more detailed time examining would exceed
the time accuracy of the data.

As one can see from the Fig.4 the empirical $C_{max}$ distribution
was shifted to the right side to meet the experimentally registered
value $C_{exp}=72.5 K^{0}$. A center of the distribution is located
close to the points $65-66 K^{0}$ that is again in a good agreement
with the forecast of gaussian approximation of data statistics
($65 K^{0}$)(!).   A new estimation of the chance probability
on respect with this distribution fits to the numbers $10^{-4}-10^{-3}$.
This estimation increases essentially the RTM value $10^{-6}$
but remains to be enough small to recognize an objectivity of the
"some correlations" between the "gravity and neutrino" data.

However having in the mind of "ill posed" character of the
problem of reconstruction of statistical distribution we have
to consider once more all sources of possible uncertainties.
The main one is of course a dependence of the result on the choice
of characteristic times which was made refering to the experiment
specifics but not to objective mathematical restriction. It is clear
that expanding a range of time shifts infinitely and shortering a step
one increases anormously a "number of occasions" and drives the
probability of chance to the unit. Thus our conclusions above have
a conventional character suspended on our argumentation for a choice
of characteristic times.

Together with this there is another, particular in the given case
but very serious, source of uncertaity which was recognized in the
process of empirical analysis. The matter is a different scale
of sampling for neutrino and gravity data: 1~s for the gravitational
detector counts but 0,01~s for accuracy of neutrino time marks.
It means that an operator has to make an {\it interpolation procedure}
trying to find a correspondent gravitational count (value of energy
innovation) to the definit neutrino event. The simplest way to do it
is a "step kind" interpolation but generally it can be done by some
optimal manner [55]. The most important point for us here is a recognition
that any interpolation introduces an uncertainty in our calculation
of the statistical parameters and in particulary in the position of the
center of $C_{max}$ distribution.

An estimation of the {\it interpolation error} [55] leads us to the
value $\Delta C_{int}=4.6 K^{0}$. This error box is shown at the
Fig. 4 around $C_{max}$ center position. A calculation gives that a
displacement of this center to the right side of the error box provides
the chance probability on the order of $0.01$ or larger. This is
a relatively typical  level for stochastic measurements and does not
mean any extraordinary event.

Thus our reanalysis shows that the available experimental
data are unsufficient to make a reliable conclusion in favour of the
detection a "remarkable $\nu g $-correlation"  during of SN1987A explosion.

\section{Discussion}

The example with SN1987A gives enough presentation how the MPL-algorithm
works exhibiting clearly at the same time its weak point:
a dependence on the unknown range of time shift  between astrophysical
and gravitational events.  An {\it apriori} estimation of it on  physical
 arguments is desirable to provide an efficiency of the algorithm.
Any attepmts to limit this range appealing to some pecularity of experimental
data or particular manner of opreator behaviour under searching  for the
"signal exitation $ C_{exp}$"  do not lead to "objective boundaries" for time
shift and thus a correspondent evaluation of the chance
probability remains to be suspended.  Only an apriori knowledge of  the
time shift range could introduce  some certainty (deterministic
elements) in our ill posed problem. In the extremely favourable case when
the value of shift is known exactly the estimation of chance probability
can be taken just from $C$-distribution which is much more robust then
 $C_{max}$-distribution.  The last one however has to be used obligatory
if the shift was not given beforehand. We can remark here that the authors
of the paper [52] have came very close to this idea introducing of so called
"$q$-parameter" to define how often a realization with rare statistical
properties occures in the process of computer simulation of experimental data.
It can be shown that such approach leads directly to  the absolute maximum
distribution.

Having in the mind a phenomenon of joint neutrino and gravitational
radiation from SN1987A  one could be limited in the range of time shift
by the theoretical restriction of the neutrino rest mass which is less then
10 ev; then a delay of neutrino signal would not exceed 2.7~s.
That is just the hypothesis which was adopted as a starting point for the
data processing in the papers [41--44], where the maximum time shift range
was taken on order of $\mp 2$ s. However due to a large uncertainty
of joint scenaria for supernova radiation dynamics [24,29,30,37] as well as
due to a general tend to avoid any hypothetical propositions concerning
a "nature of the source"  we used in our reanalysis the maximum time
shift compatible with the structure of experimental data $\mp 100$ s
The interesting fact was that even for this large time shift interval
the chance probability of  the "correlation effect" was kept on small
level $10^{-3}-10^{-4}$ and only  the "interpolation uncertainty" did not
permit to confirm a presence of the RTM-correlation.

Some alternative hypothesis for explanation of the observed experimental
data was proposed in the paper [51] where a time evolution of the
$C$-variable  on the interval of observation was presented  separately for
Roma, Maryland and combined (R+M) antennue. So the evolution
diagrams with big peak at the region of 2--4 h 23 Feb. were similary
for the combined (R+M) and Roma antennue, but  the diagram for the
Maryland antenna was different (more smooth and no big peak).
Early a correlation between seismic data and (R+M) antennae background during
of SN1987A  was reported [49].  Thus the hypothesis [51] is that it was
registered a correlation between Mont Blanc neutrino scintillator and
Roma detector backgrounds produced by a small scale  earthquake in the
south Europe region occured roughly in the SN1987A time. The  data of
Maryland detector has no evident coupling with this phenomenon.

Coming back to the general algorithm of searching for "astro-gravity
correlations" we would like make several remarks.

1) The $C$-variable in the form (19) used  in our reanalysis
is the exact MLP-variable for  a signal with unknown but
deterministic parameters: $A_{k},\Theta_{k},\tau_{k}$ . It also approximately
corresponds to the case of stochastic uniformly distributed phase;
a correct  phase averaged expression of MLP-variable in this case is given by
the formula  (18) with summation over all astrophysical events. Using (18)
one could wait a decrease of chance probability for two reasons: a) the
expression (18)  gives a more optimal estimation for small signals $A<\sigma$,
b) this is one step  from pure MLP-method to the Bayesian approach which has
in general a lower false alarm error.

2) A next step to the Bayesian aproach could be associated with an
averaging (18) also over unknown time shifts supposed  to be uniformly
distributed in the {\it apriori } given time interval.  This would produce
a following essential decrease of chance probability but the paiment will be
a refuse from evaluation of time shift between astrophysical and gravitational
data.

3) The MLP algorithm (19),(20) contains in principle a possibility
of signal accumulation. However for the "post demodulation" read out an
accumulation of small incoherent pulses $A_{k}\le \sigma$ increases
a signal-noise ratio proportionaly
to $n^{1/4}$ i.e. it can not be effective. In the opposit case
of large pulses the accumulation tends to usual low of independent
stochastic counts $n^{1/2}$ but  here it is unlikely to expect a big value
for $n$  on a reasonable observational time according to astrophysical
scenaria.

4) A search of "astro-gravity correlations" as a new
form of gravitational wave experiment has a clear advantage of
sharp reduction of  the observational time interval involved in the data
processing.  This leads to an equivalent diminution of the chance probability
proportionaly to the factor $n\Delta \tau/T$  but on a threshold
signal-noise ratio it produces a small influence increasing this ratio  only in the
$(1/2)\ln{(T/n\Delta \tau)}$ times which is insignificant.

5) We have seen in our reanalysis that the gaussian approximation gave
a good agreement with statistics found empirically. However on the wings
of empirical distributions  an uncertainty of estimations grows. A possible
way to improve a quality of empirical estimation consists in using a family of
Pirson statistics to approximate more correctly  a distribution of the
experimental data how it was proposed in [8].

6) The Pirson statistics could be also used to prognosticate of expected
probability of chance and other statistical values under a generalization
of MLP-algorithm for the case of nongaussian noises. At practice there is
always an excess of large nonthermal pulses at the tail of integral
energy distribution for gravitational bar detectors. Using the Pirson
approximation for a density probability of such nongaussian noise it is
possible to make some preliminary filtration to suppress nongaussian
hindrances. Then a generalized MLP-algorithm  will have the same form
(19),(20) with substitution of some known function
of the output overlope $f(R_{k}^{2})$ instead of overlope itself
$R_{k}^{2}$ [56].

In conclusion we would like to note that a simple translation of
the developed algorithm to the laser interferometrical antenna
on free masses is impossible without serious modification.
The matter is a response of this wide frequency band set up to
gravitational signals can not be presented in some universal form
like it was done for "quasi $\delta$-exitations" of the bar detector.
One has to take into account a complex structre of individual GW-pulses.
Thus a construction of optimal algorithm for detection of a packet
of such pulses correlated with astrophysical events becomes a
multi-parametrical problem and has to be studied specially.

\section{Aknowledgment}

The authors would like to gratitude the members of RTM group
and first of all prof.G.Pizzella,G.Pallottino and dr.S.Frasca
for a supportion of this work, providing of gravitational
experimental data and stimulative discussions. We also very
appreciated an assistance of prof.O.Ryazhskaya and S.Vernetto for
presenting neutrino event list from LSD detector and clarifing
some questions concerning its operation. Many fruitful discussions
we had with members of relativistic astrophysics group of SAI MSU
prof.K Postnov and N.Shakura. This work also was partly supported
by the Russian State Programs: "High Energy Physics"(\S 5.3/1)
and "Fundamental Nuclear Physics" (\S 1.135).

\end{document}